\documentclass[twocolumn,prl,aps,epsf]{revtex4}
\usepackage[english]{babel}
\usepackage{graphicx}
\usepackage{epsf}
\usepackage{pstricks}

\begin{document}

\title{Dispersion relations and subtractions in hard exclusive processes}

\author{I.V. Anikin, O.V. Teryaev}
\affiliation{Bogoliubov Laboratory of Theoretical Physics,
             JINR, 141980 Dubna, Russia }

\begin{abstract}

\noindent
We study analytical properties of the hard exclusive
process amplitudes. 
We found that QCD factorization 
for deeply virtual Compton scattering and hard exclusive vector meson 
production results in the subtracted dispersion 
relation with the subtraction constant determined by the Polyakov-Weiss $D$-term. 
The relation of this constant to the fixed pole contribution found by Brodsky, Close and Gunion  
and defined by 
parton distributions is studied and proved for momentum transfers exceeding the typical hadronic scale.
The continuation to the real photons limit is considered and the numerical correspondence 
between lattice simulations of $D$-term and low energy Thomson amplitude is found.    
For sufficiently large $t$ the subtraction may be expressed in the
form similar to suggested earlier for real Compton scattering.
 
\vspace{1pc}
\end{abstract}
\maketitle

\section{I. Introduction} 

Hard exclusive reactions described by the Generalized Parton 
Distributions (GPDs) \cite{DM, Ji, NonforRad, GPV, Diehl:2003ny, belrad}
are the subject of extensive theoretical and experimental studies. 
The analytical properties of deeply virtual Compton scattering (DVCS) and hard exclusive vector meson 
production (VMP) amplitudes \cite{TerPr, DMuller07, Pasq, VanderPR} constitute the important aspect of these studies. 
They play also the major role in the hadronic processes such as the nucleon-nucleon scattering  
at very high energies to be studied at LHC and in non-accelerator experiments \cite{block}.

The crucial point in application of the relevant 
dispersion relations is a possible ambiguity due to the subtraction constants which are the counterparts of 
the normalization constants implied by the ultraviolet renormalization procedure.
An attractive possibility is represented by cases when such constants are defined by the imaginary part of the 
amplitudes. This situation was explored long ago in the case of forward Compton amplitude 
\cite{Brodsky71, lee, Gerasimov:2007ec}, 
and was recently reconsidered for DVCS \cite{DMuller07}.  

In this paper we address the problem of dispersion relations and subtractions in the framework 
of the leading order QCD factorization.  
We find that it leads to subtracted 
dispersion relations with the subtraction constant defined by the Polyakov-Weiss $D$-term \cite{PW}.
At the same time, for $t$ exceeding typical hadronic scale 
we relate the subtraction constant to the integrals of parton distribution at zero skewness. 

\section{II. Dispersion relation in the skewness plane}

We restrict our study by the case of large $s$ and $Q^2$ and small $t \ll s,\, Q^2$, 
where QCD factorization is applicable.  
At the leading order, this results in the following expressions for DVCS 
and vector ($\rho^0$) meson production amplitudes: 
\begin{eqnarray}
\label{DVCSamp}
&&T_{DVCS}^{\mu\nu}=\frac{g^{\mu\nu}_\perp}{2}\bar u(p_2)\hat n u(p_1)
\sum_{f=u,d,s,...}e^2_{f}\, {\cal A}_{f}(\xi,\,t) 
\nonumber\\
\end{eqnarray}
and
\begin{eqnarray}
\label{VMPamp}
&&T_{VMP}^{\mu}=\frac{\alpha_s f_{\rho} C_F e_L^{\mu}}{\sqrt{2} N_c Q}\bar u(p_2)\hat n u(p_1)
\nonumber\\ 
&&\times{\cal V}\biggl[e_u {\cal A}_{u}(\xi,\,t)-e_d {\cal A}_{d}(\xi,\,t) \biggr] 
\end{eqnarray}
with the GPDs part (which may be interpreted as a weighted handbag diagram, i.e. the 
coupling of local quark currents to two photons)
\begin{eqnarray}
\label{ampLO}
{\cal A}_{f}(\xi,\,t)=\int\limits_{-1}^{1} dx\, 
\frac{H^{(+)}_{f}(x,\,\xi,\,t)}{x-\xi+i\epsilon}
\end{eqnarray} 
and the meson part for VMP case
\begin{eqnarray}
\label{vecpart}
{\cal V}=\int\limits_{0}^{1} dy\, \frac{\phi_1(y)}{y}.
\end{eqnarray}
In (\ref{ampLO}), $H^{(+)}(x,\,\xi,\,t)$ denotes the  singlet ($C=+1$) combination of GPDs, summing the contributions of 
quarks and anti-quarks and of $s$- and $u$-channels:
\begin{eqnarray}
H^{(+)}_{f}(x,\,\xi,\,t)=H_{f}(x,\,\xi,\,t)-H_{f}(-x,\,\xi,\,t).
\end{eqnarray}
For the sake of brevity, we will keep only  the dependence of GPDs on the skewness $\xi = Q^2/(2s+Q^2)$.
For $\xi \to 0$, (\ref{ampLO}) may acquire divergencies at $x=0$, which 
will be one of objects of our analysis.  
  
Contrary to the forward case, expression (\ref{ampLO}) does not have a form of the dispersion relation
because of the appearance of $\xi$ in the numerator. 
Nevertheless, the amplitude (\ref{ampLO}) as a function of $\xi$ manifests the analyticity in the 
unphysical region $|\xi| > 1$ \cite{TerPr}.
This region is associated with the contribution of Generalized Distribution Amplitudes (GDAs) \cite{GDA} 
related to GPDs by crossing \cite{RadonTeryaev}.
To prove the analyticity of the amplitude for $|\xi| > 1$ , one represents the denominator of (\ref{ampLO}) as 
the geometric series:
\begin{eqnarray} 
\label{decan}
{\cal A}(\xi)=- \sum_{n=0}^{\infty} \xi^{-n-1}
\int\limits_{-1}^{1} dx\,H^{(+)}(x,\,\xi) x^n.
\end{eqnarray}
This series is convergent \cite{TerPr} 
thanks to the polynomiality condition (see e.g. \cite{Diehl:2003ny, belrad}):
\begin{eqnarray}
\label{polyn}
\int\limits_{-1}^{1}dx\, x^n\, H(x,\,\xi) = 
\sum_{k=0,2...}^{n} \xi^k A_k + 
\frac{1-(-1)^n}{2}
\xi^{n+1} C.
\nonumber
\end{eqnarray}

One may now easily calculate the discontinuity across the 
cut $-1 < \xi <1$ and write the fixed-$t$ dispersion relation \cite{TerPr} for the leading order amplitude (\ref{ampLO}) 
in the skewness plane:
\begin{eqnarray}
\label{dr20}
{\rm Re}\,{\cal A}(\xi)=
\frac{{\cal P}}{2\pi i}\int\limits_{-1}^{1} dx\, 
\frac{{\rm Disc}{\cal A}(x)}{x-\xi}+\Delta(\xi).
\end{eqnarray}  
or, using (\ref{ampLO}),
\begin{eqnarray}
\label{dr2}
{\cal P}\int\limits_{-1}^{1} dx\, 
\frac{H^{(+)}(x,\,\xi)}{x-\xi}=
{\cal P}\int\limits_{-1}^{1} dx\, 
\frac{H^{(+)}(x,\,x)}{x-\xi}+\Delta(\xi),
\nonumber\\
\end{eqnarray} 
where $\Delta(\xi)$ is a possible subtraction.
This expression represents the holographic property of GPD:
the relevant information about hard exclusive amplitudes in the considered leading
approximation is contained in the one-dimensional sections $x= \pm \xi$
of the two dimensional space of $x$ and $\xi$. These holographic as well as tomographic \cite{RadonTeryaev} 
properties in momentum space 
are complementary to the often discussed  holography and tomography in coordinate space \cite{belrad}.

We are now  going to prove that $\Delta(\xi)$ is finite and independent of $\xi$, {\it i.e.} 
$\Delta(\xi)= const$.
To do this, one considers the following representation:
\begin{eqnarray} 
\label{diff2}
\Delta(\xi)={\cal P}\int\limits_{-1}^{1} dx\, 
\frac{H^{(+)}(x,\,\xi)-H^{(+)}(x,\,x)}{x-\xi} = 
\\
-{\cal P}\int\limits_{-1}^{1} dx\, \sum_{n=1}^{\infty} 
\frac{1}{n!} \frac{\partial^n}{\partial\xi^n}H^{(+)}(x,\,\xi)\biggr|_{\xi=x}(\xi-x)^{n-1}. \nonumber
\end{eqnarray}
Due to the polynomiality condition the only surviving highest power term  
in this series is equal to a finite subtraction constant. This can also be derived with a use of the
Double Distributions (DDs) formalism. Namely, the $H^{(+)}(x,\,\xi)$ is expressed through the corresponding DDs as 
\begin{eqnarray}
\label{GPDs1}
&&H^{(+)}(x,\,\xi)=\int\limits_{-1}^{1}\,d\alpha\int\limits_{-1+|\alpha|}^{1-|\alpha|}\,d\beta
\biggl[ f(\alpha,\beta) + \xi g(\alpha,\beta)\biggr] 
\nonumber\\
&&\times\biggl[ \delta(x-\alpha-\xi\beta) - \delta(-x-\alpha-\xi\beta)\biggr].
\end{eqnarray}
Substituting this expression into (\ref{diff2}), one gets 
that the $f(\alpha, \beta)$-terms which depend on $\xi$ 
are cancelled and eqn. (\ref{diff2}) becomes $\xi$-independent
(cf. \cite{TerPr}):
\begin{eqnarray}
\label{diff3}
\Delta(\xi) = -2\int\limits_{-1}^{1}\,d\alpha\int\limits_{-1+|\alpha|}^{1-|\alpha|}\,d\beta \,
\frac{g(\alpha,\beta)}{1-\beta} \equiv \Delta.
\end{eqnarray}
Notice that the cancellation of $f(\alpha,\beta)$ and validity of (\ref{diff3}) are not spoiled 
(provided $\xi \neq 0$), even if 
the singularity corresponding to $f(\alpha,\beta) \sim \alpha^{-a}$ is present. 

In (\ref{diff3}), one can choose
\begin{eqnarray}
\label{D}
g(\alpha, \beta)=\delta(\alpha)D(\beta),
\end{eqnarray}
where the function $D(\beta)$ is $D$-term \cite{PW}. 
The assumption (\ref{D}) is a result of the corresponding ``gauge'' \cite{RadonTeryaev} as discussed also in 
\cite{belrad}. With (\ref{D}), the $\Delta$ term takes the following form:
\begin{eqnarray}
\label{Deltap}
\Delta= 2 \int\limits_{-1}^{1}\,d\beta \,
\frac{D(\beta)}{\beta-1}.
\end{eqnarray} 

It should be emphasized that both integrals in (\ref{dr2}) are divergent at $\xi=t=0$, 
and this divergencies do not cancel for $\xi \to 0$. 
It means that $\Delta$ is not defined for $\xi=t=0$ \footnote{
Note that the point $\xi=0$ (or $Q^2=0$)  cannot be accessed in DVCS and VMP experiments}.
On the other hand, for an arbitrarily small $\xi$ the integrals in (\ref{dr2}) are finite and, therefore,
$\Delta$ is well-defined. 
Note some similarity between (\ref{Deltap}) and (\ref{ampLO}) so that $\Delta$ may be also 
interpreted as a contribution of local two-photon coupling to the quark currents which is independent of $\xi$. 

Taking into account the parameterization \cite{PW}
\begin{eqnarray}
\label{parD}
D(\beta)=(1-\beta^2) \sum_{n=0}^{\infty} d_n C^{(3/2)}_{2n+1}(\beta),
\end{eqnarray}
and keeping only the lowest term, one gets
\begin{eqnarray}
\label{Tmlimit2}
\Delta= - 4 d_0 .
\end{eqnarray} 
This lowest term $d_0$ was estimated within the framework of different models.
We focus on the results of chiral quark-soliton model \cite{CQM}: 
$d_0^{{\rm CQM}}(N_f)=d_0^{u}=d_0^{d}=-\frac{4.0}{N_f}$,
where $N_f$ is the number of active flavours,
and lattice simulations \cite{Lattice}: 
$d_0^{{\rm latt}}=d_0^{u}\approx d_0^{d} =-0.5$.
The subtraction constant varies as  
\begin{eqnarray}
\label{res} 
&&\Delta^{p}_{{\rm CQM}}(2)=\Delta^{n}_{{\rm CQM}}(2)\approx 4.4, 
\nonumber\\
&&\Delta^{p}_{{\rm latt}}\approx\Delta^{n}_{{\rm latt}}\approx 1.1 
\end{eqnarray}
for the DVCS on both the proton and neutron targets.

\section{III. Dispersion relation in the $\nu$ plane} 

We have seen that the $D$-term determines the 
finite subtraction in the dispersion relation in the skewness plane. Let us now compare the dispersion relation (\ref{dr2})
with the dispersion relation written in the $\nu$ plane \cite{VanderPR} where 
$\nu=(s-u)/4m_N$. 
In terms of the new variables  $\nu^\prime,\, \nu$ related to $x,\, \xi$ as 
\begin{eqnarray}
\label{rel}
x^{-1}=\frac{4m_N\nu^\prime}{Q^2}, \quad \xi^{-1}=\frac{4m_N\nu}{Q^2}, 
\end{eqnarray} 
the fixed-$t$ dispersion relation becomes the subtracted one:
\begin{eqnarray}
\label{drsub}
&&{\rm Re}\,{\cal A}(\nu, Q^2)=
\frac{\nu^2}{\pi}{\cal P}\int\limits_{\nu_0}^{\infty} \frac{d\nu^{\prime\, 2}}{\nu^{\prime\,2}}\,  
\frac{{\rm Im}\,{\cal A}(\nu^{\prime}, Q^2)}{(\nu^{\prime\, 2}-\nu^2)}+\Delta=
\nonumber\\
&&\frac{{\cal P}}{\pi}\int\limits_{\nu_0}^{\infty} d\nu^{\prime\, 2}
{\rm Im}\,{\cal A}(\nu^{\prime}, Q^2)\biggl[ \frac{1}{\nu^{\prime\, 2}-\nu^2} -  
\frac{1}{\nu^{\prime\, 2}}\biggr]+ \Delta.
\end{eqnarray}
Here, $\nu_0=Q^2/4m_N$ (the $Q^2$ dependence here is shown explicitly) and 
the nucleon pole term residing in this point may be considered separately \cite{VanderPR}.

This subtracted (in the symmetric unphysical point $\nu=0$)
dispersion relation is the principal result of our paper.
It is applicable for both DVCS (cf. \cite{Pasq, VanderPR}) and VMP (cf. \cite{BrodEstr}) amplitudes. 

It can be considerably simplified provided ${\rm Im}\,{\cal A}(\nu)$ decreases fast enough 
so that both terms in the squared brackets can be integrated separately:  
\begin{eqnarray}
\label{dr1}
{\rm Re}\,{\cal A}(\nu)=
\frac{{\cal P}}{\pi}\int\limits_{\nu_0}^{\infty} d\nu^{\prime\, 2}\, 
\frac{{\rm Im}\,{\cal A}(\nu^{\prime})}{\nu^{\prime\, 2}-\nu^2} + {\bf C}_0 \, ,
\end{eqnarray}
where 
\begin{eqnarray}
\label{C0}
{\bf C}_0&=& \Delta -
\frac{{\cal P}}{\pi}\int\limits_{\nu_0}^{\infty} d\nu^{\prime\, 2}\,  
\frac{{\rm Im}\,{\cal A}(\nu^{\prime})}{\nu^{\prime\,2}}
\nonumber\\
& =& \Delta + {\cal P}\int\limits_{-1}^{1} dx\, 
\frac{H^{(+)}(x,\,x)}{x}.
\end{eqnarray}
Now, using (\ref{diff2}) with $\xi=0$, one gets: 
\begin{eqnarray}
\label{subtr1}
\Delta &=& {\cal P}\int\limits_{-1}^{1} dx\, 
\frac{H^{(+)}(x,\,0)-H^{(+)}(x,\,x)}{x}
\\
\label{subtr12}
&=&2{\cal P}\int\limits_{-1}^{1} dx\, 
\frac{H(x,\,0)-H(x,\,x)}{x},
\end{eqnarray}
where the symmetry property arising from the $T$-invariance:
$H(x,-x)=H(x,x)$ is used. 
The relation (\ref{subtr1}) can also be obtained from the ``sum rules'' \cite{GPV}:
\begin{eqnarray}
\label{polyn}
\int\limits_{-1}^{1}dx\frac{H(x,\xi+xz)- H(x,\xi)}{x} = 
\sum_{n=1}^{\infty}z^n
\int\limits_{-1}^{1}dxx^{n-1}D(x)
\nonumber
\end{eqnarray}   
for $\xi=0$ and $z=1$.

Let us stress that for the valence ($C=-1$) contributions to the amplitudes of the hard exclusive 
production of, say, pions \cite{GPV} and exotic
hybrid mesons \cite{AnHybrid}, $\Delta=0$ because of the mentioned symmetry in $x$ and the $\xi$-independence.

Substituting (\ref{subtr1}) into (\ref{C0}), one can see that the $D$-term is cancelled from the expression for 
the subtraction constant
\begin{eqnarray}
\label{Const}
{\bf C}_0(t)
= 2{\cal P}\int\limits_{-1}^{1} dx 
\frac{H(x,\,0,\,t)}{x},
\end{eqnarray}
where we restored the dependence on $t$ which was omitted for brevity.
This constant is similar to the result obtained in the studies of the fixed pole 
contribution to the forward Compton amplitude \cite{Brodsky71}.
At the same time, at $\xi,\, t=0$, GPDs are expressed in terms of standard parton distributions 
$H(x,\,0)=q(x)\theta(x)-\bar q(-x)\theta(-x)$. Formally one has 
\begin{eqnarray}
{\bf C}_0(0)&=&2\int\limits_{0}^{1} dx 
\frac{q(x)+\bar q(x)}{x} 
\nonumber\\
&=& 2\int\limits_{0}^{1} dx 
\frac{q_v(x)+ 2 \bar q(x)}{x}.
\end{eqnarray}
 
However, the integral defining ${\bf C}_0(0)$ diverges at low $x$ in both the valence and sea quark contributions.
Therefore, for $t=0$ we should consider (\ref{drsub}) as a correct general form of the dispersion relation which includes
the infinite subtraction at the point $\nu=0$ and the subtraction constant associated with the $D$-term.  

For $t\neq 0$, the integral in (\ref{Const}) converges for sufficiently large $t$. 
In the case of Regge inspired parameterization \cite{GPV} $H(x,\,0,\,-t)\sim x^{-\alpha(0)+\alpha^\prime t}$,  
this condition reads as $t>\alpha(0)/\alpha^\prime$,
resulting in $t \gtrsim 1 (10){\rm GeV}^{2}$ for the valence (sea)
quark distributions.

The divergence of (\ref{Const}) was originally discussed within the framework of the parton model
and its modifications \cite{Brodsky71} while we address this problem in the framework of the leading order QCD 
factorization. Therefore, our result (\ref{drsub}), although being formally $\xi$ independent, cannot, generally speaking, 
be  continued to the forward limit $\xi\sim Q^2\to 0 \, (s=const)$. 
The limit $\xi\to 0, \, s\to\infty,\, Q^2=const$
still corresponds to the highly non-forward kinematics with the masses of initial and final photons 
being rather different.  
At the same time, further exploration of the possible manifestation of 
the $D$-term in the forward Compton scattering seems very interesting. 

The formal continuation of (\ref{drsub}) would result in the following subtracted dispersion 
relation for the forward Compton scattering amplitude:
\begin{eqnarray}
\label{drsubF}
&&{\rm Re}\,{\cal A}(\nu)=
\frac{\nu^2}{\pi}{\cal P}\int\limits_{0}^{\infty} \frac{d\nu^{\prime\, 2}}{\nu^{\prime\, 2}}\,  
\frac{{\rm Im}\,{\cal A}(\nu^{\prime})}{(\nu^{\prime\,2}-\nu^2)} + \Delta.
\end{eqnarray}

Comparing this expression with the dispersion relation for the forward Compton scattering
amplitude \cite{Damashek}, one can observe an interesting numerical coincidence.
For the proton target, our subtraction combined with the 
lattice simulations (\ref{res})
is rather close to the low energy Thomson term (note that $\Delta_{{\rm Thomson}}=1$ for our normalization of the 
Compton amplitude).
As a result, the mysterious occurrence of Thomson term at large energies \cite{Damashek} can now be supplemented 
with its possible appearance also at large $Q^2$.  
This does not hold in the case of neutron target where 
DVCS subtraction term is the same, while Thomson term is zero. This may be because 
the sum of squares of valence quark charges is equal to the square of proton 
charge, {\it i.e.} to the square of their sum (cf. \cite{CI}),  whereas for the neutron these quantities differ.

\section{IV. Conclusions} 

In this paper we show that the fixed-$t$ dispersion relations 
for the DVCS and VMP amplitudes  
require the infinite subtractions at the unphysical point $\nu=0$ with the subtraction 
constants associated with the $D$-terms. 
However, for the productions of the mesons defined by valence ($C=-1$) GPDs
the finite subtraction is absent.  
  
We also show that the appearance of the subtraction expressed in terms of (forward) parton distributions 
\cite{Brodsky71} 
may be investigated in the framework of the leading order QCD factorization. 
We consider the possibility of continuation of our results to the real photons limit. The 
surprising similarity between the lattice simulations of the $D$-term 
and the low energy Thomson  amplitude in the proton target case is found.

\section{Acknowledgments} 

We would like to thank A.P.~Bakulev, M.~Diehl, 
 A.V.~Efremov, S.B.~Gerasimov, D.Yu.~Ivanov, D.~ M\"uller, B.~Pire, M.V.~Polyakov, A.V.~Radyushkin, L.~Szymanowski, 
S.~Wallon for useful discussions and correspondence.  
This work was supported in part by Deutsche Forschungsgemeinschaft
(Grant 436 RUS 113/881/0), RFBR (Grants 06-02-16215 and 07-02-91557) and 
RF MSE RNP (Grant 2.2.2.2.6546).

\paragraph{Note added.}-- After this work was completed, the papers \cite{DI,P07} appeared confirming our 
results and generalizing them to the next-to-leading order.  



\begin{thebibliography}{99}
\vspace{1\baselineskip}

\bibitem{DM}
  D.~Mueller, D.~Robaschik, B.~Geyer, F.~M.~Dittes and J.~Horejsi,
  Fortsch.\ Phys.\  {\bf 42}, 101 (1994)
  [arXiv:hep-ph/9812448].

\bibitem{Ji}
  X.~D.~Ji,
  Phys.\ Rev.\  D {\bf 55}, 7114 (1997)
  [arXiv:hep-ph/9609381].

\bibitem{NonforRad}
  A.~V.~Radyushkin,
  Phys.\ Rev.\  D {\bf 56}, 5524 (1997)
  [arXiv:hep-ph/9704207].

\bibitem{GPV}
  K.~Goeke, M.~V.~Polyakov and M.~Vanderhaeghen,
  Prog.\ Part.\ Nucl.\ Phys.\  {\bf 47}, 401 (2001)
  [arXiv:hep-ph/0106012].

\bibitem{Diehl:2003ny}
  M.~Diehl,
  Phys.\ Rept.\  {\bf 388}, 41 (2003)
  [arXiv:hep-ph/0307382].

\bibitem{belrad}
  A.~V.~Belitsky and A.~V.~Radyushkin,
  Phys.\ Rept.\  {\bf 418}, 1 (2005)
  [arXiv:hep-ph/0504030].


\bibitem{Pasq}
  B.~Pasquini, M.~Gorchtein, D.~Drechsel, A.~Metz and M.~Vanderhaeghen,
  Eur.\ Phys.\ J.\  A {\bf 11}, 185 (2001)
  [arXiv:hep-ph/0102335].

\bibitem{VanderPR}
  D.~Drechsel, B.~Pasquini and M.~Vanderhaeghen,
  Phys.\ Rept.\  {\bf 378}, 99 (2003)
  [arXiv:hep-ph/0212124].

\bibitem{TerPr}
  O.~V.~Teryaev, ``Analytic properties of hard exclusive amplitudes,''
  in proc. of 
  11th Intern. Conf. on Elastic and Diffractive Scattering 
  (Chateau de Blois, May 2005), 
  ed. by  M. Haguenauer, B.Nicolescu, J. T$\hat{\rm r}$an 
  Thanh V$\hat{\rm a}$n, p. 237 
  [arXiv:hep-ph/0510031].
  
\bibitem{DMuller07}
  K.~Kumericki, D.~Muller and K.~Passek-Kumericki,
  arXiv:hep-ph/0703179.

\bibitem{block}
M.~M.~Block,
  Phys.\ Rept.\  {\bf 436}, 71 (2006)
  [arXiv:hep-ph/0606215].

\bibitem{Brodsky71}
  S.~J.~Brodsky, F.~E.~Close and J.~F.~Gunion,
  Phys.\ Rev.\  D {\bf 5}, 1384 (1972); {\it ibid.}
   D {\bf 8}, 3678 (1973).
  \bibitem{lee}
  S.~Y.~Lee,
  Nucl.\ Phys.\  B {\bf 45}, 449 (1972).

\bibitem{Gerasimov:2007ec}
  S.~B.~Gerasimov,
  arXiv:hep-ph/0701073.

\bibitem{PW}
  M.~V.~Polyakov and C.~Weiss,
  Phys.\ Rev.\  D {\bf 60}, 114017 (1999)
  [arXiv:hep-ph/9902451].

\bibitem{GDA}
  M.~Diehl, T.~Gousset, B.~Pire and O.~Teryaev,
  Phys.\ Rev.\ Lett.\  {\bf 81}, 1782 (1998)
  [arXiv:hep-ph/9805380].

\bibitem{RadonTeryaev}
  O.~V.~Teryaev,
  Phys.\ Lett.\  B {\bf 510}, 125 (2001)
  [arXiv:hep-ph/0102303].

\bibitem{CQM}
  V.~Y.~Petrov, P.~V.~Pobylitsa, M.~V.~Polyakov, I.~Bornig, K.~Goeke and C.~Weiss,
  Phys.\ Rev.\  D {\bf 57}, 4325 (1998)
  [arXiv:hep-ph/9710270].

\bibitem{Lattice}
  M.~Gockeler, R.~Horsley, D.~Pleiter, P.~E.~L.~Rakow, A.~Schafer, G.~Schierholz and W.~Schroers
                  [QCDSF Collaboration],
  Phys.\ Rev.\ Lett.\  {\bf 92}, 042002 (2004)
  [arXiv:hep-ph/0304249].

\bibitem{BrodEstr}
  S.~J.~Brodsky and F.~J.~Llanes-Estrada,
  Eur.\ Phys.\ J.\  C {\bf 46}, 751 (2006)
  [arXiv:hep-ph/0512247];
  M.~Vanderhaeghen, P.~A.~M.~Guichon and M.~Guidal,
  Phys.\ Rev.\  D {\bf 60}, 094017 (1999)
  [arXiv:hep-ph/9905372].

\bibitem{AnHybrid}
  I.~V.~Anikin, B.~Pire, L.~Szymanowski, O.~V.~Teryaev and S.~Wallon,
  Phys.\ Rev.\  D {\bf 70}, 011501 (2004)
  [arXiv:hep-ph/0401130];
  Phys.\ Rev.\ D {\bf 71}, 034021 (2005)
  [arXiv:hep-ph/0411407];
  Nucl.\ Phys.\  A {\bf 755}, 561 (2005)
  [arXiv:hep-ph/0501119];
  arXiv:hep-ph/0509245.

\bibitem{Damashek}
  M.~Damashek and F.~J.~Gilman,
  Phys.\ Rev.\  D {\bf 1}, 1319 (1970).
  
\bibitem{CI}
  F.~E.~Close and N.~Isgur,
  Phys.\ Lett.\  B {\bf 509}, 81 (2001)
  [arXiv:hep-ph/0102067].
    
\bibitem{DI}
  M.~Diehl and D.~Y.~Ivanov,
  arXiv:0707.0351 [hep-ph].

\bibitem{P07}
M.~V.~Polyakov,
  arXiv:0707.2509 [hep-ph].

\end{thebibliography}
\end{document}